\documentclass{ws-procs9x6-cpt22}
\begin{document}

\newcommand{\refeq}[1]{(\ref{#1})}
\def\etal {{\it et al.}}
\def\al{\alpha}
\def\be{\beta}
\def\ga{\gamma}
\def\de{\delta}
\def\ep{\epsilon}
\def\ve{\varepsilon}
\def\ze{\zeta}
\def\et{\eta}
\def\th{\theta}
\def\vt{\vartheta}
\def\io{\iota}
\def\ka{\kappa}
\def\la{\lambda}
\def\vpi{\varpi}
\def\rh{\rho}
\def\vr{\varrho}
\def\si{\sigma}
\def\vs{\varsigma}
\def\ta{\tau}
\def\up{\upsilon}
\def\ph{\phi}
\def\vp{\varphi}
\def\ch{\chi}
\def\ps{\psi}
\def\om{\omega}
\def\Ga{\Gamma}
\def\De{\Delta}
\def\Th{\Theta}
\def\La{\Lambda}
\def\Si{\Sigma}
\def\Up{\Upsilon}
\def\Ph{\Phi}
\def\Ps{\Psi}
\def\Om{\Omega}
\def\cA{{\cal A}}
\def\cB{{\cal B}}
\def\cC{{\cal C}}
\def\cE{{\cal E}}
\def\cF{{\cal F}}
\def\cl{{\cal L}}
\def\cL{{\cal L}}
\def\cO{{\cal O}}
\def\cP{{\cal P}}
\def\cR{{\cal R}}  
\def\cV{{\cal V}}  
\def\cS{\Si}
\def\cT{{\cal T}}

\def\mn{{\mu\nu}}
\def\abgd{{\al\be\ga\de}}

\def\fr#1#2{{{#1} \over {#2}}}
\def\half{{\textstyle{1\over 2}}}
\def\quar{{\textstyle{1\over 4}}}

\def\vev#1{\langle {#1}\rangle}
\def\bra#1{\langle{#1}|}
\def\ket#1{|{#1}\rangle}
\def\bracket#1#2{\langle{#1}|{#2}\rangle}
\def\expect#1{\langle{#1}\rangle}
\def\norm#1{\left\|{#1}\right\|}
\def\abs#1{\left|{#1}\right|}

\def\lsim{\mathrel{\rlap{\lower4pt\hbox{\hskip1pt$\sim$}}
    \raise1pt\hbox{$<$}}}
\def\gsim{\mathrel{\rlap{\lower4pt\hbox{\hskip1pt$\sim$}}
    \raise1pt\hbox{$>$}}}
\def\sqr#1#2{{\vcenter{\vbox{\hrule height.#2pt
         \hbox{\vrule width.#2pt height#1pt \kern#1pt
         \vrule width.#2pt}
         \hrule height.#2pt}}}}
\def\square{\mathchoice\sqr66\sqr66\sqr{2.1}3\sqr{1.5}3}

\def\prt{\partial}
\def\lrpartial{\raise 1pt\hbox{$\stackrel\leftrightarrow\partial$}}
\def\lrprt{\stackrel{\leftrightarrow}{\partial}}
\def\lrprtnu{\stackrel{\leftrightarrow}{\partial^\nu}}
\def\lrDmu{\stackrel{\leftrightarrow}{D_\mu}}
\def\lrDnu{\stackrel{\leftrightarrow}{D^\nu}}
\def\lrvec#1{ \stackrel{\leftrightarrow}{#1} }
\def\part2{\partial_\alpha \partial^\alpha}

\def\Re{\hbox{Re}\,}
\def\Im{\hbox{Im}\,}
\def\Arg{\hbox{Arg}\,}

\def\hb{\hbar}
\def\etal{{\it et al.}}

\def\pt#1{\phantom{#1}}
\def\ni{\noindent}
\def\ol#1{\overline{#1}}

\def\ss{$s^{\mu\nu}$}
\def\tt{$t^{\ka\la\mu\nu}$}
\def\uu{$u$}

\def\sss{s^{\mu\nu}}
\def\ttt{t^{\ka\la\mu\nu}}

\def\sb{\overline{s}}
\def\tb{\overline{t}}
\def\ub{\overline{u}}

\def\stw{\tilde{s}}
\def\ttw{\tilde{t}}
\def\utw{\tilde{u}}
\def\Btw{\tilde{B}}

\def\hsy{h_{\mu\nu}}  
\def\nsy{\et_{\mu\nu}}

\def\he{h^E}
\def\hp{h'}

\def\xx'{|\vec x -\vec x'|}

\def\lrDmu{{\hskip -3 pt}\stackrel{\leftrightarrow}{D_\mu}{\hskip -2pt}}

\def\nsc#1#2#3{\om_{#1}^{{\pt{#1}}#2#3}}
\def\lsc#1#2#3{\om_{#1#2#3}}
\def\usc#1#2#3{\om^{#1#2#3}}
\def\lulsc#1#2#3{\om_{#1\pt{#2}#3}^{{\pt{#1}}#2}}

\def\tor#1#2#3{T^{#1}_{{\pt{#1}}#2#3}}

\def\vb#1#2{e_{#1}^{{\pt{#1}}#2}}
\def\ivb#1#2{e^{#1}_{{\pt{#1}}#2}}
\def\uvb#1#2{e^{#1#2}}
\def\lvb#1#2{e_{#1#2}}

\def\disp#1{\displaystyle{#1}}

\def\b2{b^\al b_\al}
\def\ff{\ve}

\newcommand{\beq}{\begin{equation}}
\newcommand{\eeq}{\end{equation}}
\newcommand{\bea}{\begin{eqnarray}}
\newcommand{\eea}{\end{eqnarray}}
\newcommand{\bit}{\begin{itemize}}
\newcommand{\eit}{\end{itemize}}
\newcommand{\rf}[1]{(\ref{#1})}
\newcommand\bw{\begin{widetext}}
\newcommand\ew{\end{widetext}}

\title{A Four-Parameter Black-Hole Solution\\ in the Bumblebee Gravity Model}

\author{Rui Xu}

\address{Kavli Institute for Astronomy and
Astrophysics, Peking University, Beijing 100871, China}

\begin{abstract}
The bumblebee gravity model includes a class of vector-tensor theories of gravitation where the vector field couples to the Ricci tensor quadratically. We obtain an analytical spherical black-hole solution in this model. The solution has four parameters, expanding the two-parameter solution family known in the literature. Special choices of the parameters are pointed out and discussed. 
\end{abstract}

\bodymatter

\section{The bumblebee gravity model}
The bumblebee gravity model is given by the action\cite{ref1}
\bea
S &=& \int d^4x \sqrt{-g} \left( \frac{1}{2\ka} R + \frac{\xi}{2\ka} B^\mu B^\nu R_\mn - \frac{1}{4} B^\mn B_\mn - V \right), 
\label{actionB}
\eea  
where $\ka=8\pi$,\footnote{We use the geometrized unit system where the gravitational constant and the speed of light are set to 1.} $R$ and $R_\mn$ are the Ricci scalar and Ricci tensor, respectively, and $B_\mu$ is the bumblebee vector field. The constant $\xi$ controls the size of the coupling between the bumblebee field and the Ricci tensor. The field strength of $B_\mu$ is $B_\mn = D_\mu B_\nu - D_\nu B_\mu$ with $D_\mu$ being the covariant derivative. The potential $V$ is introduced to set up the nonzero background of the bumblebee field, namely that $V$ is minimized when the bumblebee field takes some nonzero background configuration $B_\mu = b_\mu$. When this happens, in a local Lorentz frame the coupling term $b^ab^bR_{ab}$ violates local Lorentz invariance as the local background configuration $b^a$ does not change under a local Lorentz transformation by definition.\footnote{Here we mean a particle local Lorentz transformation; see Ref.~[\refcite{ref1}] for details.}

The bumblebee model has been studied intensively in the SME framework, where the background configuration of the bumblebee vector field $b_\mu$ is usually assumed to be constant in a Minkowski background spacetime but allowing both the spacetime metric and the bumblebee field to have small fluctuations around their background values.\cite{ref1,ref2,ref3,ref4,ref5,ref6} One intriguing result is that the fluctuation of the bumblebee field (not the full bumblebee field) might be interpreted as the EM vector potential, leaving the possibility that photons are actually products of Lorentz-invariance violation!\cite{ref2} The background value of the bumblebee field then corresponds to coefficients for Lorentz violation in the SME, and is constrained by high-precision experiments aiming at testing Lorentz symmetry.\cite{refa1}

Outside the SME framework, solutions to the bumblebee model in the strong-field regime, i.e., black-hole solutions, have also been investigated.\cite{ref7,ref8,ref9,ref10} In this case, the background bumblebee field depends on the spacetime coordinates. A background bumblebee field with only the radial component was found for a Schwarzschild-like metric in Ref.~[\refcite{ref8}]. Inspired by it, we generalize the solution with a nontrivial temporal component of the background bumblebee field in this work.

\section{Finding the analytical solution}\label{aba:sec1}
When considering the background configuration that minimizes the potential $V$ in Eq.~\rf{actionB}, the potential does not contribute when taking variations with respect to the fields. So the field equations take the form 
\bea
&& G_\mn = \left( T_{b}\right)_\mn , 
\nonumber \\
&& D^\mu b_{\mn} + 2 \xi b^\mu R_\mn  = 0,
\label{fieldeqs2}
\eea
where we have used the background bumblebee field $b_\mu$ in the equations, and its energy--momentum tensor is
\bea
\left( T_{b}\right)_\mn &=& \frac{\xi}{2} \Big[ g_\mn b^\al b^\be R_{\al\be} - 2 b_\mu b_\la R_\nu^{\pt\nu \la} - 2 b_\nu b_\la R_\mu^{\pt\mu \la} - \Box_g ( b_\mu b_\nu ) 
\nonumber \\
&& - g_{\mn} D_\al D_\be ( b^\al b^\be ) + D_\ka D_\mu \left( b^\ka b_\nu \right) + D_\ka D_\nu ( b_\mu b^\ka )   \Big]
\nonumber \\
&& + \ka \left[ b_{\mu\la} b_\nu^{\pt\nu \la} - \frac{1}{4} g_\mn  b^{\al\be} b_{\al\be} \right]  .
\eea
We start by investigating the vector-field equation for $b_\mu = \left( b_t(r),\, b_r(r),\, 0,\, 0\right)$ with the Schwarzschild metric. 
Surprisingly, the $r$-component of the vector field equation vanishes automatically, leaving the $t$-component of the equation
\bea
\frac{d^2b_t}{dr^2}  + \frac{2}{r} \frac{db_t}{dr} = 0, 
\eea
which gives
\bea
b_t = C + \frac{D}{r},
\label{btsol}
\eea
with $C$ and $D$ being integration constants. The simple expression made us to suspect that $(T_b)_\mn$ can be zero with a properly specified $b_r$. In fact, after substituting Eq.~\rf{btsol} into $(T_b)_\mn$, we find $(T_b)_\mn$ is indeed zero as long as 
\bea
b_r^2 = \frac{ 6\xi (C M + D)CM r + \xi D^2 (2r-M) - \ka D^2 (r-2 M) }{ 3 \xi M (r-2 M)^2 } ,
\eea
where $M$ is the mass of the Schwarzschild black hole.

Next, it is straightforward to verify that the Schwarzschild-like metric
\bea
ds^2 = -\left( 1 - \frac{2M}{r} \right)dt^2 + \frac{1+l}{1-\frac{2M}{r}} dr^2 + r^2 \left(d\th^2 + \sin^2\th d\ph^2\right),
\label{metricsol}
\eea
together with Eq.~\rf{btsol} and 
\bea
b_r^2 &=& \frac{ (1+l)\left[6\xi (C M + D)CM r + \xi D^2 (2r-M) - \ka D^2 (r-2 M)\right] }{ 3 \xi M (r-2 M)^2 } 
\nonumber \\
&& + \frac{l(1+l)}{\xi} \frac{r}{r-2M} ,
\label{brsol}
\eea
satisfies the field equations \rf{fieldeqs2}, where $l,\,M,\,C$, and $D$ are four parameters that characterize the solution.

\section{Discussion}
By setting $C=D=0$, the solution given by Eqs.~\rf{metricsol}, \rf{btsol}, and \rf{brsol} recovers the solution in Ref.~[\refcite{ref8}]. With $l=0$ but nonzero $C$ and $D$, we have a more interesting situation now: the Schwarzschild metric is accompanied by a nontrivial background bumblebee field. We especially would like to draw the readers' attention to the case of $\xi=2\ka$. In this specific bumblebee model, if $D=-2MC$ then $b_t \propto 1 - 2M/r$ and $b_r=0$. 

It is also inspiring to consider the parameter choice $l=0$ and $M=0$ so that the metric becomes the Minkowski metric. Though the $t$-component of the background bumblebee field trivially becomes constant, $b_r$ has a nontrivial limit if $D \propto \sqrt{M}\rightarrow 0$. As the background bumblebee field in the Minkowski spacetime breaks Lorentz symmetry, and if we believe the broken of Lorentz symmetry originates from a fundamental theory of quantum gravity, then the ratio $D^2/M$, representing the length scale of the background bumblebee field, is likely to be several Planck lengths.

Current solar-system observations set stringent constraints on the parameter $l$ as shown in Ref.~[\refcite{ref8}]. The newly introduced parameters $C$ and $D$ do not make any difference when considering test particles moving along geodesics. However, the existence of the background bumblebee field certainly affects motions of binary black holes as well as GWs emitted. Also, if the fluctuation of the bumblebee field is identified as the EM vector potential, then we expect the trajectory of light to deviate from null geodesics, leaving imprints in the shadow images of black holes. It is therefore worthwhile to develop perturbation solutions in the background metric and the background bumblebee field shown here. 
As GW data from coalescences of binary black holes accumulate and the resolution of the Event Horizon Telescope improves,\cite{ref11,ref12,ref13,ref14,ref15} unprecedented tests are on the way and they may eventually distinctively favor or disprove the bumblebee model.



\end{document}